\documentclass[11pt]{article}

\usepackage[margin=1in]{geometry}
\usepackage{amsmath,amssymb}
\usepackage{xcolor}
\usepackage{graphicx} 
\usepackage{tikz}
\usepackage{pdfpages}
\usetikzlibrary{arrows.meta, positioning, backgrounds, calc}
\usepackage{xspace}
\usepackage{microtype}
\usepackage{indentfirst}
\usepackage[numbers,sort&compress]{natbib}   
\usepackage[colorlinks=true,linkcolor=blue!55!black,citecolor=blue!55!black,
            urlcolor=blue!55!black]{hyperref}
\usepackage{listings}
\lstdefinestyle{src}{
  basicstyle=\ttfamily\scriptsize,
  breaklines=true,
  breakatwhitespace=false,
  columns=fullflexible,
  keepspaces=true,
  showstringspaces=false,
  extendedchars=true,
  frame=single,
  framesep=4pt,
  xleftmargin=2pt,
  aboveskip=6pt,belowskip=6pt,
  tabsize=2,
}
\newcommand{\code}[1]{\texttt{\small #1}}
\newcommand{\tool}[1]{\texttt{#1}}
\usepackage{booktabs}
\newcommand{\jj}{\ensuremath{J_1 - J_2}\xspace}

\newcommand{\sysname}{\textsc{QuantumMatterExplorer}\xspace}   

\title{\sysname: An Autonomous Agent for the Discovery of Quantum Many-Body Phases of Matter}

\usepackage{authblk}

\usepackage{orcidlink}

\author[1]{Jannes Nys\,\orcidlink{0000-0001-7491-3660}}
\author[1]{Andrew Jreissaty\,\orcidlink{0009-0004-2539-3127}}
\author[1]{Juan Carrasquilla\,\orcidlink{0000-0001-7263-3462}}

\affil[1]{Institute for Theoretical Physics, ETH Zürich, 8093 Zürich, Switzerland}
\affil[ ]{\small Corresponding author:
\href{mailto:jcarrasquill@ethz.ch}{\texttt{jcarrasquill@ethz.ch}}}

\date{\today}

\begin{document}
\maketitle

\begin{abstract}
Mapping the phase diagram of a new quantum Hamiltonian demands expert judgement at every step: which observables to measure, where in parameter space to compute, or how to separate genuine order from finite-size artefacts. We introduce \sysname, an autonomous agent that takes only a Hamiltonian and its parameter ranges as input and returns a phase diagram obtained from its own numerical simulations. The agent is built on a large language model and is designed to reason over a long horizon, maintain a persistent written plan, and drive a set of computational tools, with density-matrix renormalization group (DMRG) calculations cross-checked against independent exact diagonalization. The agent combines a large language model with persistent planning, a library of expert computational-physics skills, and density-matrix renormalization group calculations validated against independent exact diagonalization. It adaptively selects quantities to evaluate, refines parameter scans, and performs finite-size analyses as evidence accumulates. After validating the approach on Hamiltonians with known transitions, we challenge it with a frustrated bilinear-biquadratic spin-1 chain with single-ion anisotropy, whose two-dimensional phase diagram was published only recently and is unlikely to appear in any training corpus. Given no hint about the expected physics, the agent proposes which observables to probe, and identifies candidate N\'eel, Haldane and dimerized phases, in broad agreement with the reference phase diagram. It also detects incommensurate correlations and infers the frustration-dependent geometry of the phase boundaries, all within an eight-hour compute budget. These results open a practical route to the autonomous computational discovery of new phases of quantum matter.
\end{abstract}

\tableofcontents

\section{Introduction}

Over the last few decades, much of our understanding of quantum many-body phenomena has been shaped by computational methods. Methods such as exact diagonalization, quantum Monte Carlo, and density-functional theory have unravelled new phase diagrams and exposed emergent behavior in correlated regimes inaccessible by purely theoretical analysis~\cite{foulkes2001qmc,jones2015dft}. Arguably the most powerful method today that has transformed our understanding of strongly correlated quantum systems is the density-matrix renormalization group (DMRG), introduced by White in 1992~\cite{white1992density, fannes1992finitely}. Together with its generalization to tensor networks~\cite{verstraete2004renormalizationalgorithmsquantummanybody}, DMRG has become the method of choice for one- and quasi-one-dimensional lattice models and forms a reliable tool for discovering new physical phenomena and phases of matter. It has underpinned much of our understanding of quantum magnetism, Hubbard and $t$--$J$ physics, spin liquids, and non-equilibrium dynamics~\cite{schollwock2011,orus2014}.

Three decades of development have turned DMRG into a systematic and remarkably reliable tool. Its accuracy is mainly controlled by a single key parameter: the bond dimension. Furthermore, a large number of open-source implementations are accessible, such as ITensor~\cite{itensor} (and many others including TeNPy, quimb, Block2, Cytnx, and TensorKit~\cite{itensor,hauschild2018tenpy,gray2018quimb,zhai2023block2,wu2025cytnx,devos2025tensorkit,sehlstedt2025software}). This maturity makes DMRG ideal for automated discovery, provided it can be successfully paired with intelligent interpretations, together with expert knowledge on how to explore phase diagrams or identify new quantum phenomena. 

In the last few years, large language models (LLMs) have gone far beyond chatbots, and have been paired with tools to create agents for task automation~\cite{wang2024agentsurvey, xi2025agentsurvey}. In this paradigm, LLMs are embedded in a loop where they reason, call tools, observe the results, and act again. An increasing amount of research explores the automation of various aspects of research, from hypothesis generation to code execution and manuscript writing, giving rise to so-called ``AI (co-)scientists'' to streamline scientific research~\cite{ren2025survey,zhou2025autonomous}. Given the reliability and stability of DMRG and exact diagonalization (ED), in this work we introduce \sysname: a tool for automated phase diagram exploration of new quantum Hamiltonians.  In other words, our tool automatically explores the physics of unseen Hamiltonians through computational exploration, which aims to go beyond simply reproducing known phase diagrams or human-generated analyses.

Our work contributes the following:
\begin{itemize}
    \item We demonstrate autonomous phase-diagram exploration of a likely unseen quantum Hamiltonian, extending recent work on autonomous quantum simulation~\cite{li2026autonomous,deng2026verifiable} and the exploration of initially unknown physical systems~\cite{naegele2025sciexplorer}. Given the Hamiltonian, our agent autonomously selects the observables and order parameters used to identify candidate phases and determines where in parameter space to perform calculations. It combines DMRG~\cite{white1992density,itensor} with ED~\cite{xdiag}, allowing numerical results to be cross-validated across independent methods.
    \item The exploration is adaptive rather than based on a predetermined computational protocol: the agent uses intermediate numerical results to decide what to investigate next, refining parameter-space cuts, system sizes and diagnostics as evidence accumulates.
    \item Because standard language models lack specialized training data on numerical phase exploration, we explicitly equip our agent with expert domain knowledge. We quantify the impact of this added expertise through a controlled ablation study.
    \item We compare multiple language-model backbones, including open-weight models served locally, under an identical protocol, and report their accuracy in locating known phase transitions.
\end{itemize}

\section{Related work}\label{sec:relatedwork}

We group the related literature into (i)~general autonomous AI scientists, and (ii)~LLM agents for quantum many-body systems and across the broader quantum sciences.

\subsection{Autonomous AI scientists}\label{sec:autonomousAIscientists}

While the idea of automating discovery predates LLMs (e.g.,\ the robot scientists Adam and Eve, which combined hypothesis generation with physical lab automation to autonomously discover functional-genomics knowledge~\cite{king2009automation}), the latter have revived interest and expanded the possibilities. Sakana's AI Scientist explored end-to-end automation for machine-learning research~\cite{lu2024aiscientist, lu2026nature}, including ideation, experimentation, plotting, paper writing, and automated review. The second version produced what the authors describe as the first fully AI-generated paper to pass peer review at a scientific workshop~\cite{yamada2025aiscientist2}. Other approaches explored variations to the pipeline, including orchestrating a team of role-playing agents (PhD/postdoc/professor) acting as research assistants~\cite{schmidgall2025agentlab}, among other configurations~\cite{schmidgall2025agentrxiv, gottweis2025coscientist}. In the physical sciences, some noteworthy variations include \textsc{cmbagent}, an agentic planning-and-control system that carried out a PhD-level cosmology analysis~\cite{laverick2024cmbagent}, and the \textsc{Denario} project, which targets research spanning multiple disciplines~\cite{villaescusa2025denario}. For recent in-depth reviews, we refer to Refs.~\cite{zhou2025autonomous, ding2026verification}.

The most human part of research is arguably idea generation. Nevertheless, various works have explored the possibility of automating this aspect of the process as well. Notably, Gu and Krenn generate and rank ideas that are validated by human experts~\cite{gu2024ideas}, and their AI-Mandel prototype couples arXiv-driven idea pools with dedicated novelty- and feasibility-scoring agents~\cite{arlt2025autonomous}. Other works have further explored idea generation, aiming to improve the novelty and diversity of the proposals~\cite{li2024coi, hu2024nova, zhao2025deepideation}. An important recurring finding is that adversarial agentic interactions (e.g.,\ an idea generator paired with a skeptical critic) yield stronger proposals than single-agent prompting~\cite{villaescusa2025denario}.

\subsection{Agentic systems for quantum many-body simulation}

LLM agents are being introduced throughout the quantum sciences. For example, in quantum chemistry, \textsc{El Agente} dynamically generates, executes, and debugs quantum-chemistry workflows from natural-language prompts using a hierarchy of specialized agents, achieving high degrees of success on university-level exercises~\cite{zou2025elagente, perezsanchez2026quntur}. Similarly, Ref.~\cite{zhang2026forjador} describes a system that generates task-specific agents to automate quantum simulation. In quantum computing, multi-agent systems can produce OpenQASM circuit code~\cite{fu2025qagent}, and LLM-driven systems autonomously design variational quantum circuits~\cite{sakka2026llm}. They have also proven valuable in quantum-computing laboratories, where agents autonomously calibrate superconducting processors~\cite{xu2026vibe,cao2024kagents} and prepare and characterize entangled states~\cite{cao2024kagents}.

Agentic systems have also been explored to automate parts of the workflow for running quantum many-body computations. Examples include frameworks for automating scientific simulations across different software packages and disciplines~\cite{meng2026fermilink}, density-functional-theory calculations for materials research~\cite{wang2025dreams}, and validation-driven workflows for atomistic modeling on supercomputers~\cite{dawson2026lara}. Another stream of research is based on the idea introduced by \textsc{SciExplorer}, aiming to discover physics from observations and measurements~\cite{naegele2025sciexplorer}. The agent recovers equations of motion from observed dynamics as well as Hamiltonians from expectation values. It explores initially unknown physical systems without any domain-specific blueprint, using a minimal set of tools built around code execution. 

For strongly correlated matter relevant to, e.g., condensed matter physics, Ref.~\cite{li2026autonomous} introduces multi-agent systems to autonomously write and run tensor-network (DMRG) simulations, and to produce a number of requested observables and corresponding visualizations. To achieve this, the authors explore in-context learning by providing the agents with curated documentation of DMRG simulations. In a similar vein, Ref.~\cite{deng2026verifiable} aims to build multi-agent systems that reproduce the results of analyses from high-impact journals. They open-sourced this effort by releasing QMP-Bench, a benchmark dataset of around 100 computational quantum physics tasks. Both approaches are, by construction, reproduction tasks, meaning the target phase diagram or figure is known in advance. This is in stark contrast to the current work, where we aim for open-ended discovery using computational techniques as an exploration tool. Complementary to our work, on the experimental side, phase diagram exploration is also being automated by using language models to steer high-throughput experimentation~\cite{tamura2026phase}.

\section{Methodology}\label{sec:methodology}

\begin{figure}[tbh]
    \centering
    \includegraphics[width=0.95\linewidth]{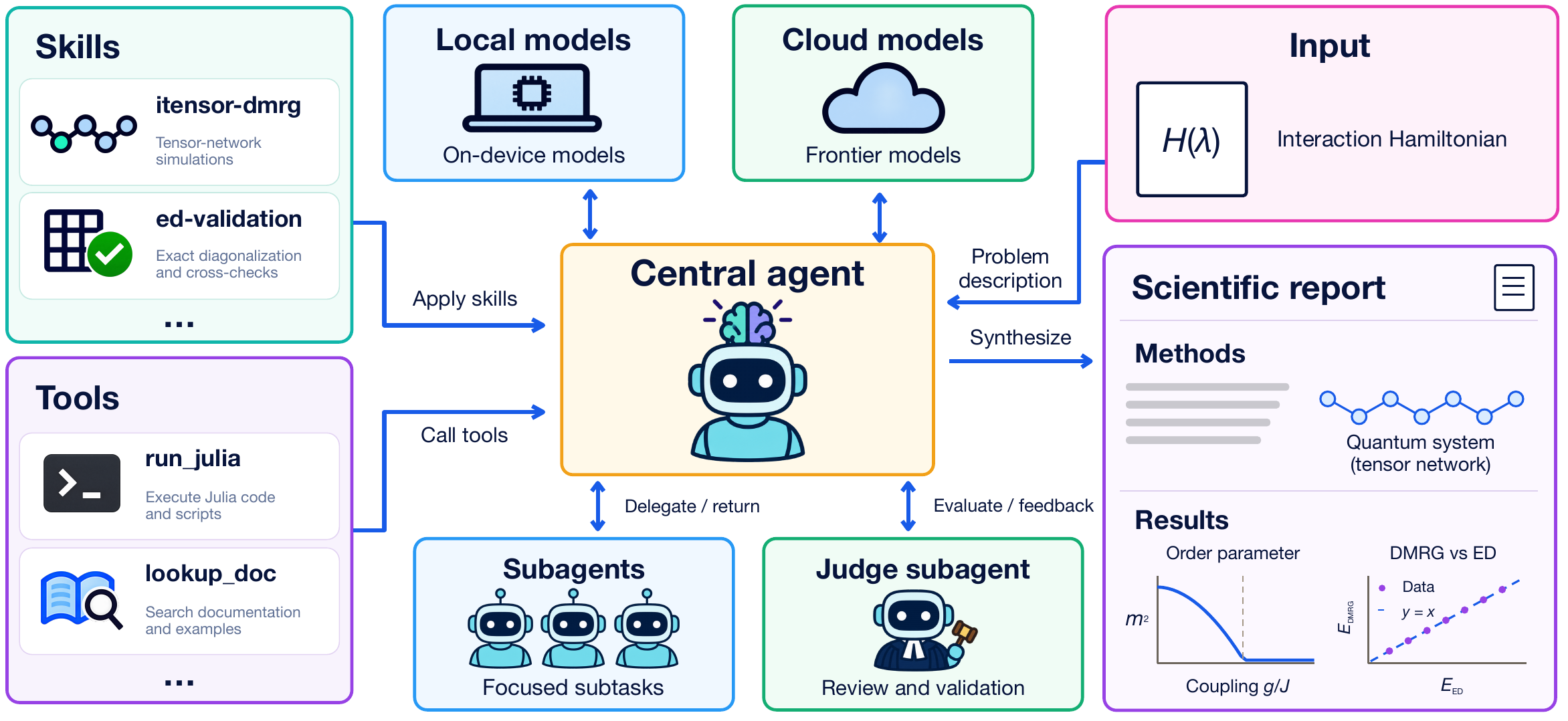}
    \caption{Overview of the \sysname architecture.}
    \label{fig:overview}
\end{figure}

Mapping a quantum phase diagram is an open-ended, multi-step research task. A practitioner must: choose observables and order parameters that identify the nature of a phase, write correct code, survive the crashes and convergence failures encountered during exploration, interpret numerical output, adjust and adapt the plan in observe-decide-rerun loops, perform finite-size scaling to separate true order from finite-size artefacts, and document the whole study so it is reproducible. We therefore take the act-and-observe loop of ReAct~\cite{yao2023react} as the basic building block of the tool. An agentic approach in a loop with tools, a filesystem, and a way to compress its own history (referred to as ``context compaction''), addresses these challenges. The overall goal of our method is to obtain a full description of a phase diagram, given the description of a quantum spin Hamiltonian (see Appendix~\ref{sec:input_example} for an example). We do so by developing a strategy to execute numerical simulations. An overview of the architecture of \sysname is given in Fig.~\ref{fig:overview}.

\paragraph{Agent harness and subagents} Whether a task benefits from being split across several agents remains an empirical question. Recent work, however, points to recurring failure modes of inherently multi-agent systems, most notably the loss or distortion of information as it passes between agents~\cite{cemri2025mast}, and finds that their reported advantages largely vanish at matched token budgets~\cite{tran2026singleagent} or come at up to ten times the cost of a plain chain-of-thought baseline~\cite{jwalapuram2026illusion}. In this work we therefore use a single ``deep agent''~\cite{deepagents} that is given access to a large number of tools. This mirrors the default configuration of popular tools such as Claude Code~\cite{anthropic_claude_code_2026}. In this approach, one central reasoner holds the entire trajectory of the study. At the same time, the agent can spawn subagents (via the \tool{task} tool), which are invoked for bounded subtasks, work in isolated contexts, and return compact results to the main agent, the latter retaining the plan and the decision history throughout. What we avoid is the alternative in which the study is partitioned across many specialised agents wired into a state machine~\cite{wu2023autogen,qian2023chatdev}, so that each agent sees only the fields the orchestrator forwards and no single agent holds the full picture. In preliminary experiments with that arrangement, we observed over-looping, redundant file rewriting, repeated exploration of the same region of the phase diagram, and decisions taken on partial information, in line with the information-loss failures reported in Ref.~\cite{cemri2025mast}. 

Subagents therefore serve here as a mechanism for context offloading rather than for the division of authority~\cite{lin2026contextenv}. We use three types. (1) A general-purpose subagent to which the main agent hands a narrowly specified subtask. (2) The \tool{data-analyst}, which reads and summarizes large data files and runs numerical post-processing, so that bulk data never enters the main context. (3) The \tool{physics-critic}, an independent, skeptical reviewer of the plan and the conclusions. Adding a skeptical critic has been found to yield stronger output than single-agent prompting~\cite{villaescusa2025denario}, and we combine LLM-as-a-judge evaluation~\cite{zheng2024judge} with tool-based validation~\cite{gou2024critic}. Importantly, its isolated context keeps it from being biased by the original main agent's reasoning. The physics critic is typically called to review the initial plan (\code{plan.md}, see below), and at the end of analysis phases to critique the current observations and simulations.

\paragraph{Flow and planning}  Two ingredients are needed for long-horizon tasks that last several hours. First, the information produced in such runs exceeds any context window, and the management of context has become a subject in its own right~\cite{lodha2026lesscontext,lin2026contextenv,verma2026activecompression}. We rely on the automatic LLM-based compaction of the history mentioned above. Second, explicit planning improves long-horizon behavior: separating a planner from an executor raises success rates on long web-navigation tasks~\cite{erdogan2025planandact}, and a controlled study of programming agents finds that supplying a written plan improves performance and that periodically reminding the agent of it strengthens adherence~\cite{liu2026planfollow}. 

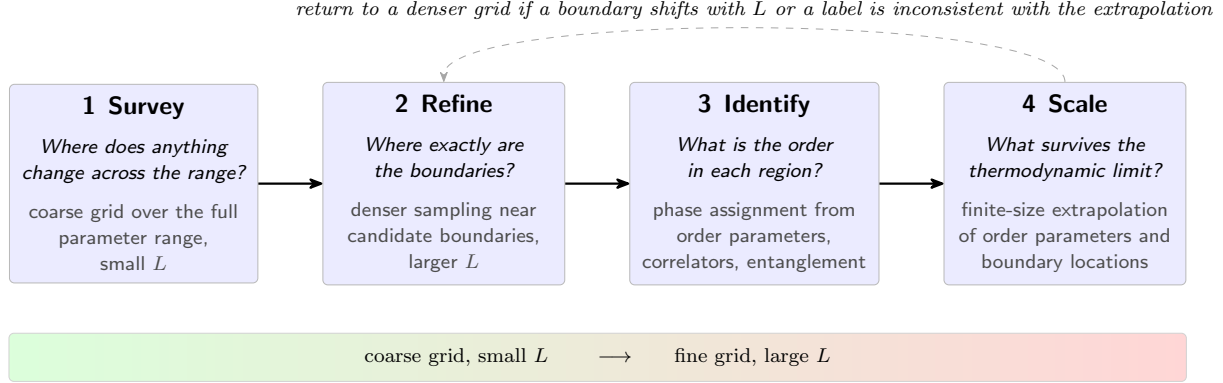
\begin{figure}[tbh]
  \centering
  \resizebox{0.98\linewidth}{!}{\begin{tikzpicture}[
  font=\sffamily\small,
  >={Stealth[round]},
  box/.style={draw=gray!55,rounded corners=2pt,align=center,inner sep=5pt,
              minimum width=34mm,minimum height=21mm},
  ph/.style={box,fill=blue!8},
]
\newcommand{\q}[1]{{\scriptsize\itshape #1}}
\newcommand{\m}[1]{{\scriptsize\color{black!70} #1}}

\node[ph] (survey)
  {\textbf{1\;\,Survey}\\[3.5pt]
   \q{Where does anything}\\[-1.5pt]\q{change across the range?}\\[3.5pt]
   \m{coarse grid over the full}\\[-1.5pt]\m{parameter range,}\\[-1.5pt]\m{small $L$}};

\node[ph, right=9mm of survey] (refine)
  {\textbf{2\;\,Refine}\\[3.5pt]
   \q{Where exactly are}\\[-1.5pt]\q{the boundaries?}\\[3.5pt]
   \m{denser sampling near}\\[-1.5pt]\m{candidate boundaries,}\\[-1.5pt]\m{larger $L$}};

\node[ph, right=9mm of refine] (identify)
  {\textbf{3\;\,Identify}\\[3.5pt]
   \q{What is the order}\\[-1.5pt]\q{in each region?}\\[3.5pt]
   \m{phase assignment from}\\[-1.5pt]\m{order parameters,}\\[-1.5pt]\m{correlators, entanglement}};

\node[ph, right=9mm of identify] (scale)
  {\textbf{4\;\,Scale}\\[3.5pt]
   \q{What survives the}\\[-1.5pt]\q{thermodynamic limit?}\\[3.5pt]
   \m{finite-size extrapolation}\\[-1.5pt]\m{of order parameters and}\\[-1.5pt]\m{boundary locations}};

\draw[->,line width=0.9pt] (survey)   -- (refine);
\draw[->,line width=0.9pt] (refine)   -- (identify);
\draw[->,line width=0.9pt] (identify) -- (scale);

\draw[->,dashed,gray!70] (scale.north) .. controls +(0,10mm) and +(0,10mm) .. (refine.north)
  node[midway,above,font=\scriptsize\itshape,text=black]
  {return to a denser grid if a boundary shifts with $L$ or a label is inconsistent with the extrapolation};

\begin{scope}[on background layer]
  \shade[left color=green!14,right color=red!16,draw=gray!35,rounded corners=2pt]
    ([yshift=-7mm]survey.south west) rectangle ([yshift=-13.5mm]scale.south east);
\end{scope}
\node[font=\scriptsize] at
  ($([yshift=-10.25mm]survey.south west)!0.5!([yshift=-10.25mm]scale.south east)$)
  {coarse grid, small $L$ \qquad$\longrightarrow$\qquad fine grid, large $L$};
\end{tikzpicture}}
  \caption{The exploration protocol. As a first step (not shown), an independent validation gate certifies the Hamiltonian and observables implementation against known limits and an exact-diagonalization cross-check before any result is trusted. A cheap coarse survey at the smallest system size then localizes candidate boundaries (Stage~1), so that the expensive large-$L$ computation is concentrated only in windows around them (Stages~2--4) rather than spent uniformly over the parameter range. Each region is then characterized by its finite-size behavior, by searching for a possible order parameter, the excitation gap, or a topological signature that survives the $L\to\infty$ limit (Stage~3). Each transition is characterized by extrapolation $L\to\infty$ (Stage~4). The flow is result-dependent, and a region whose data disagree is revisited.}
  \label{fig:phases}
\end{figure}

At the start of the process, an elaborate plan is written down by \sysname (to \code{plan.md}).  The written plan survives automatic context compaction~\cite{verma2026activecompression}, and hence limits the two dominant long-horizon failure modes: the loss of earlier constraints after compression and the compounding of an early unverified assumption~\cite{wang2026longhorizonmirage}. The core of this plan consists of four stages of exploration suggested in the system prompt. We start with (0) a first validation step to reproduce a known limit and cross-check the implementation of the Hamiltonian against independent exact diagonalization, which directly addresses the verification gap identified in Ref.~\cite{ding2026verification}. As sketched in Fig.~\ref{fig:phases}, we then follow four steps of numerical exploration: (1)~a coarse survey across the control parameters, (2)~a refined boundary detection from the survey, (3)~identification of the order of the phases and properties of the phase boundaries, and (4)~upscaling and finite-size scaling to accurately locate transitions and extrapolate to the thermodynamic limit. This decomposition is standard practice in computational condensed matter physics. We further improve and extend the prompts and skills through iterative prompt engineering using Claude Opus 5~\cite{anthropic2026opus5}. Apart from the overall plan, a separate document maintains an up-to-date description of all the computational simulations that have been carried out, similar to a lab notebook used by researchers.

\paragraph{Skills} The capabilities of the LLM are extended with skills, which serve as a method for (human) expert knowledge to enter the agent's knowledge base. This forms a training-free alternative to post-training the model for tool use on a niche library~\cite{cui2026agenticrl}, and the mechanism by which a single agent can take over the role of a team of specialised ones~\cite{li2026skills}. The current skill set includes (1) guidance on different stages of how to perform a phase diagram exploration, and (2) code-library-related skills that avoid common pitfalls in niche software packages. For the coding skills (2), we currently include compressed information about the ITensors, ITensorMPS, and XDiag libraries, which have significantly improved the performance of the agent, especially when using smaller and open LLM models (such as Gemma~4~\cite{gemma4}). A small description of the skills is always available in context, but can be read in detail when deemed necessary or useful for the ongoing analysis. The library is deliberately kept small, well below the size at which skill selection has been observed to degrade~\cite{li2026skills}. The full list of skills is summarized in Table~\ref{tab:skills}.

\paragraph{Tools} The agent further has access to multiple tools that are described in its system prompt. For a full overview, we refer to Table~\ref{tab:tools}. Note that since proper planning is key to a long-term run, we further encourage loops: plan $\to$ execute $\to$ analyse $\to$ replan. Aside from the human-readable plan, we therefore give the agent access to an internal to-do list, since, as noted above, re-presenting the plan during execution strengthens adherence to it~\cite{liu2026planfollow}. This further facilitates the planning of more detailed concrete substeps, together with fixes or changes to apply later on. Furthermore, the agent can ask for human feedback, which can edit the human-readable plan and suggest changes. This Human-In-The-Loop (HITL) approach is not used in our experiments below, but is useful in practice.

\sysname is agnostic to the language-model backend and requires only that the model supports tool calling. We consider three open-weight models served locally with vLLM~\cite{kwon2023vllm}, Gemma~4-31B-it~\cite{gemma4}, Qwen3.6-27B~\cite{qwen36}, and gpt-oss-120b~\cite{gptoss2025} (117B parameters in a mixture of experts architecture, 5.1B active per token), as well as two cloud-hosted models, DeepSeek-V4-Pro~\cite{deepseekv4} and GPT-5.6-Luna~\cite{openai2026gpt56}.

\section{Results}\label{sec:results}
\begin{figure}[htb]
    \centering
    \includegraphics[width=0.9\linewidth]{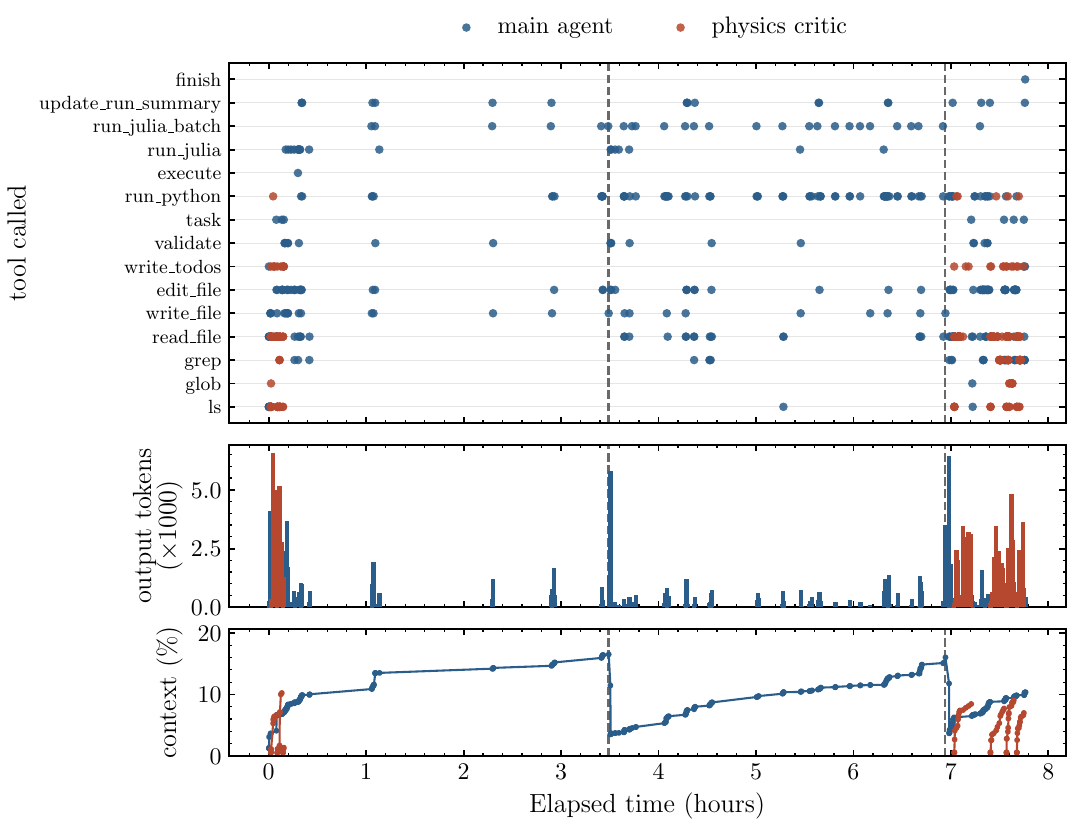}
    \caption{
    \textbf{Trajectory with tool calls of one autonomous study} using GPT-5.6-Luna at high reasoning effort on the spin-1 chain of Eq.~\eqref{eq:blbq}. (A) Composition of the 653 tool calls and 453 model calls throughout the trajectory. Colors indicate the agent type performing the tool calls. (B) Output tokens per LLM call. (C) Fraction of the context window in use per agent type. Every subagent has its own context, which can be observed in the drop in context usage. The dashed lines mark the moments a context compression is triggered, which we impose at a fixed input token budget. The total study has a maximum time limit of 8 hours, which the agent anticipates, finalizing the execution at 7.7 hours.
    \label{fig:flow}}
\end{figure}

We first demonstrate a typical trajectory of the analysis, where LLMs interact with the various available tools in a ReAct-type flow to generate simulations and analyze results in an iterative loop. Figure~\ref{fig:flow} shows such a progression for the spin-1 chain of Eq.~\eqref{eq:blbq}, and demonstrates that the tool goes through various stages similar to a computational physicist: plan, implement, validate, compute, analyze, write.

The first few minutes are spent on reading skills and managing the output directory, together with writing to-dos and planning (\tool{ls}, \tool{glob} and \tool{read\_file}). An implementation in Julia then follows (\tool{write\_file}, \tool{edit\_file}, \tool{validate}), with the first DMRG calls already at around $0.2$ h. The DMRG analysis occupies the bulk of the run, where batched Hamiltonian parameter scans (\tool{run\_julia\_batch}) alternate with visualization and data analysis in Python (\tool{run\_python}) in a measure-then-interpret loop that continues until $7$ h. A critic is spawned early (near $0.1$-$0.3$ h) to verify the plan, and then repeatedly over the final $30$ minutes to check the results and conclusions, which is visible as the dense red band at the right of all three panels. Two context compressions are triggered, near $3.5$ h and $6.9$ h, and the context window is forced to stay below $20\%$ throughout to reduce the context size.

\begin{figure}[htb]
    \centering
    \includegraphics[width=0.8\linewidth]{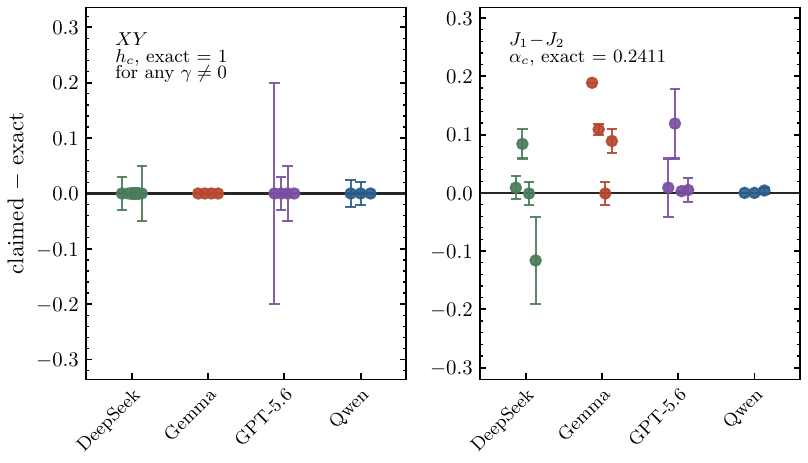}
    \caption{\textbf{Accuracy of the reported transition points relative to their reference values}. We show the numerically estimated transition points for the anisotropic XY and \jj models across multiple agentic runs using four LLM backends, following hyperparameter tuning (including LLM temperature).  Error bars indicate the uncertainties stated by the agents in their respective research notes; no error bar is shown when no explicit uncertainty was provided. The XY critical field $h_c = 1$ is exact (free-fermion solution), whereas the \jj dimerization point $\alpha_c \approx 0.2411$ is determined numerically in Ref.~\cite{okamoto1992}.
    }
    \label{fig:fig2b_transition_accuracy}
\end{figure}

\subsection{Validation}

While the cost of computation is a bottleneck in computational studies, the quality of the analysis is largely determined by the expertise in planning and interpretation. This consists in selecting, among others, which observables to measure and which system sizes to consider, as well as devising guidance on how to interpret the results. We evaluate \sysname along three different axes: accuracy, reproducibility, and cost/behavior.

A first challenge in judging the accuracy of \sysname on the open-ended task of phase exploration is that its output consists of scientific conclusions, many of which may be up for debate. Nevertheless, to score the performance, we first choose two benchmark Hamiltonians whose phase diagrams are known, such that conclusions that are obtained numerically can be validated against a known ground truth. We therefore consider the anisotropic XY chain in a transverse magnetic field [Eq.~\eqref{eq:xy}], which admits an exact free-fermion solution throughout its parameter space, and the frustrated \jj chain [Eq.~\eqref{eq:j1j2}], whose phase diagram features the exactly solvable Majumdar--Ghosh point, with a ground state given by a product of nearest-neighbor singlets \cite{majumdar1969} (see Appendix~\ref{sec:hamiltonians} for details).

As mentioned, \sysname can operate with any LLM backend that allows for tool interaction. We compare the performance of various models. In Fig.~\ref{fig:fig2b_transition_accuracy}, we visualize how accurately the numerical results produced by the agent pinpoint a known phase boundary. We observe that all agents reliably reproduce the phase transition of the XY model, but struggle more to locate the Berezinskii–Kosterlitz–Thouless (BKT) transition between the gapless and dimerized phases of the \jj chain \cite{okamoto1992}. Of all the selected backends, we find that Qwen3.6-27B performs the strongest here, despite it being an open and small model. In particular, it appears to combine strong coding capabilities with excellent exploration and analysis skills.

\subsection{Ablation}
\begin{figure}[htb]
    \centering
    \includegraphics[width=0.6\linewidth]{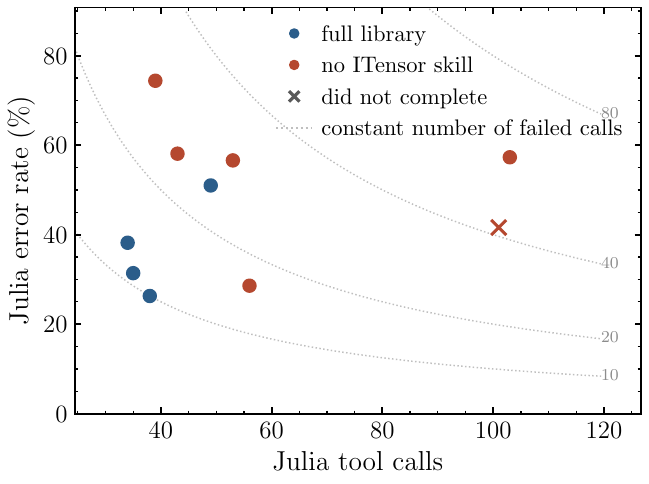}
    \caption{\textbf{ITensor skills ablation study}. Julia error rate versus the total number of Julia calls for multiple independent runs on the XY Hamiltonian, with (blue) and without (red) the \tool{itensor-dmrg} skill available. The red cross indicates an experiment that reaches the 8-hour time budget due to a large number of failed coding attempts. Dashed lines show a constant number of Julia failures (failed = total calls $\times$ failure rate). The results are shown for a local Gemma 4 model.}
    \label{fig:fig9d_julia_skill_effort}
\end{figure}

Running DMRG calculations with a specialised software tool such as ITensor is known to be challenging for LLM-based agents~\cite{deng2026verifiable}, partly due to the limited representation of such tools in their training data~\cite{li2026autonomous}. This is reflected in the relatively high frequency of errors encountered when agents generate and run DMRG code with ITensor. We observe that including the \tool{itensor-dmrg} skill significantly reduces both the agent's Julia error rate and the total number of Julia calls, as well as the absolute number of failed calls, for a Gemma 4 backend on a relatively straightforward Hamiltonian such as the XY model, as demonstrated in the ablation study in Fig.~\ref{fig:fig9d_julia_skill_effort}. In addition, we ran the same study with a different but larger model, gpt-oss-120b (117B MoE, 5.1B active), served locally under the same protocol, and observed that it failed to produce a working DMRG pipeline at either medium or maximum reasoning effort, even with the \tool{itensor-dmrg} skill available.

\subsection{Cost}
\begin{figure}[htb]
    \centering
    \includegraphics[width=1.0\linewidth]{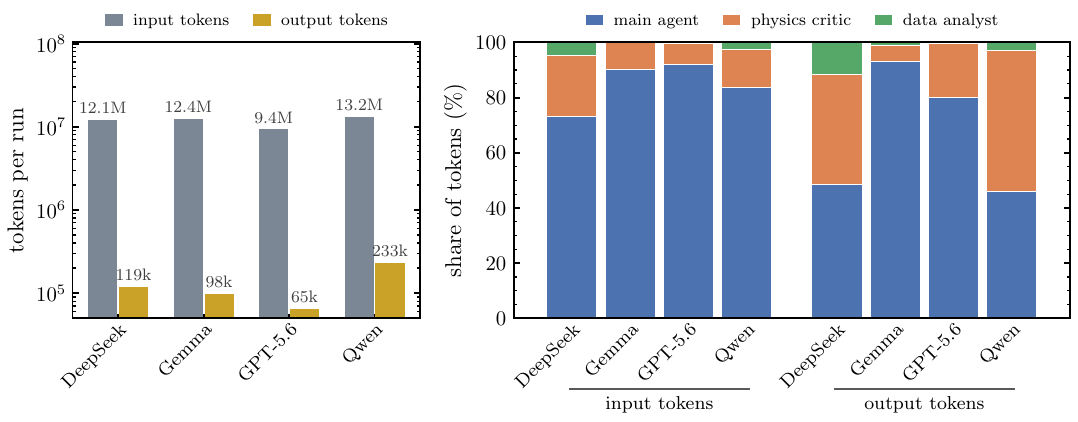}
    \caption{\textbf{Token usage}. (left) Mean token usage per run for each backbone. (right) The share of each agent type (main, critic, or data analyst). The results are averaged over 2 Hamiltonians (XY and $\jj$), with 4 repeated runs each. }
    \label{fig:fig7_agent_split}
\end{figure}

We consider the cost of \sysname in terms of LLM usage. Figure~\ref{fig:fig7_agent_split} shows the total input and output token usage for the various LLM backends used in this study, together with a breakdown across the three agent types. We observe that input tokens substantially outnumber output tokens, with all backbones consuming similar input budgets, while Qwen3.6 tends to generate considerably more output. Its research notes also tend to provide a more accessible description of the phase diagram. For cloud-based inference, output tokens are in general also significantly more expensive than input tokens. We additionally report the wall-clock time of the simulations, broken down into LLM inference time and tool usage, in Appendix~\ref{sec:timeconsumption} (see Fig.~\ref{fig:fig7b_time_split}), together with an estimate of the cost of cloud-based inference.

\subsection{Exploring an undocumented phase diagram}\label{sec:blbq}

\begin{figure}[tbh]
    \centering
    \includegraphics[width=0.5\linewidth]{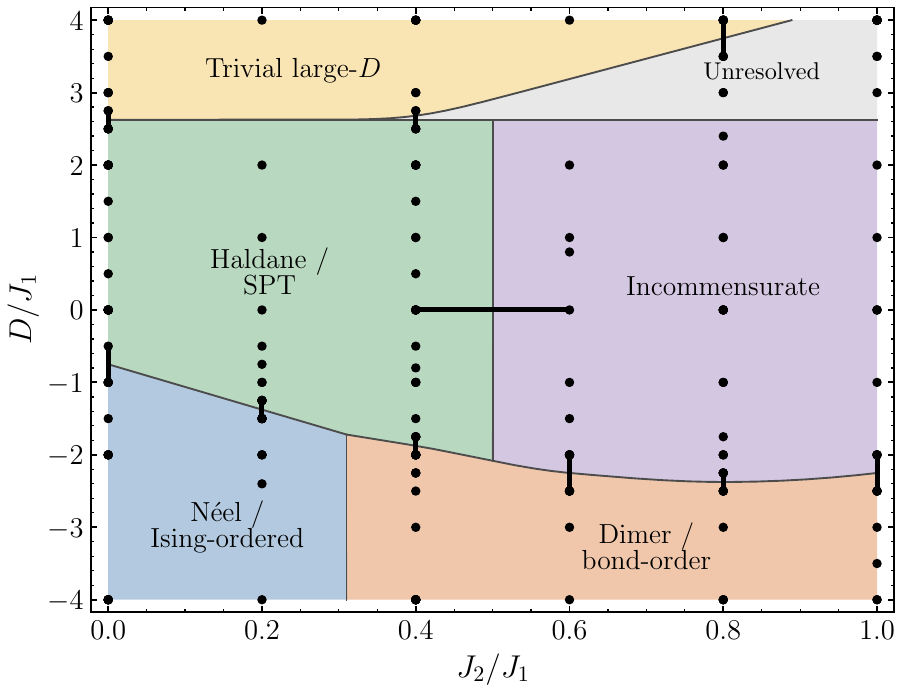}%
    \includegraphics[width=0.5\linewidth]{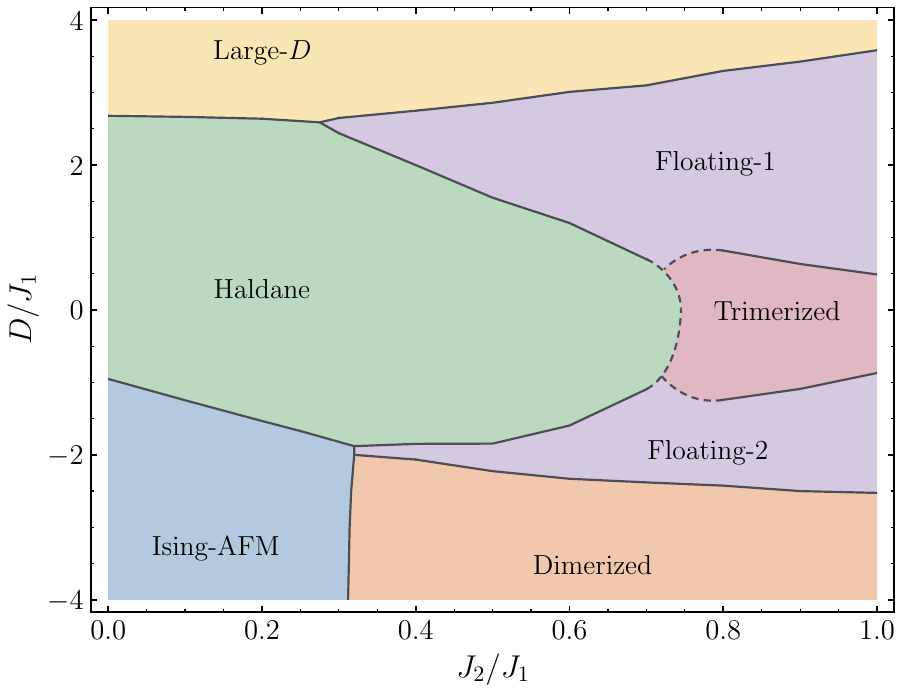}%

    \caption{\textbf{The phase diagram produced for Eq.~\eqref{eq:blbq_main_body} at $\theta=0.2\pi$.} (left) A rough sketch of the five candidate regions identified by the agent. Black markers indicate Hamiltonian parameter points computed with DMRG, while thick bars indicate finite-size phase-boundary brackets reported by the agent. The agent assigns candidate phases to only a subset of the computed points; the colored regions and their boundaries are our interpolation of these assignments and the reported brackets, rather than point-by-point classifications or thermodynamic transition estimates. The unresolved region contains points for which the agent could not confidently assign a phase. (right) Adapted phase diagram from Ref.~\cite{lariviere2026}, with dashed lines indicating uncertainty in the locations of the phase transitions. Details concerning the nature of the transition lines in Ref.~\cite{lariviere2026} are omitted for clarity.}
    \label{fig:blbq_phase_diagram_compare}
\end{figure}

\begin{figure}[tbh]
    \centering
    \includegraphics[width=0.6\linewidth]{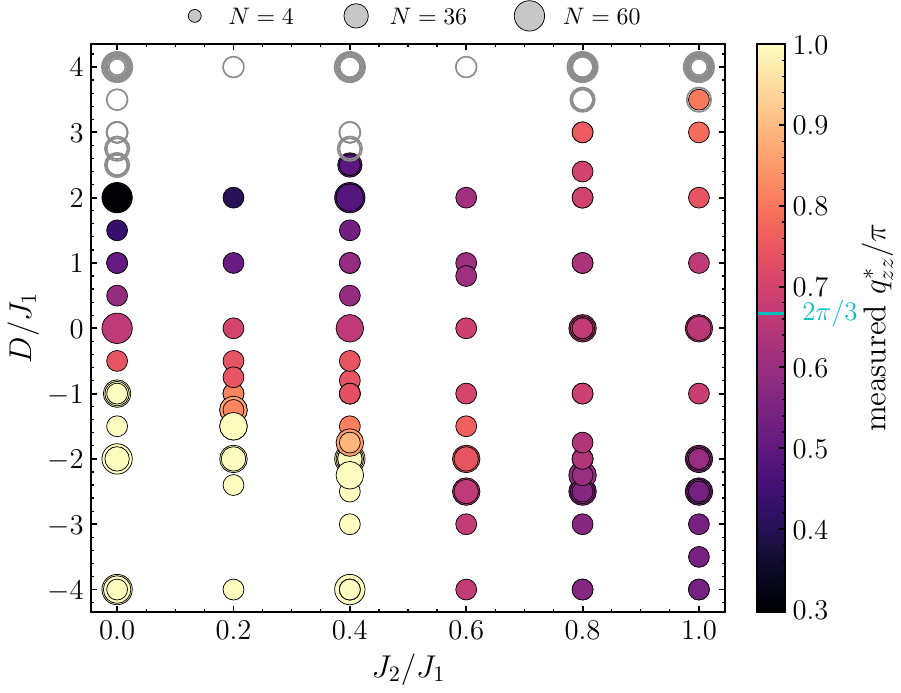}%
    \caption{\textbf{Longitudinal spin-correlation wavevector for Eq.~\eqref{eq:blbq} at $\theta=0.2\pi$.} The measured ordering wavevector $q^*$ that maximizes Eq.~\eqref{eq:structure_factor} with $\alpha=z$ at the computed parameter points, with marker size proportional to the chain length ($N\leq60$). Open grey markers denote points where $S_{zz}(q)$ is almost flat. The wavevector is $q^*=\pi$ in the N\'eel-ordered corner and moves continuously through the frustrated regime, reaching $0.665\pi \approx 2\pi/3$ (marked on the color bar) along $D=0$.}
    \label{fig:blbq_phase_diagram_points}
\end{figure}

\begin{figure}[tbh]
    \centering
    \includegraphics[width=0.95\linewidth]{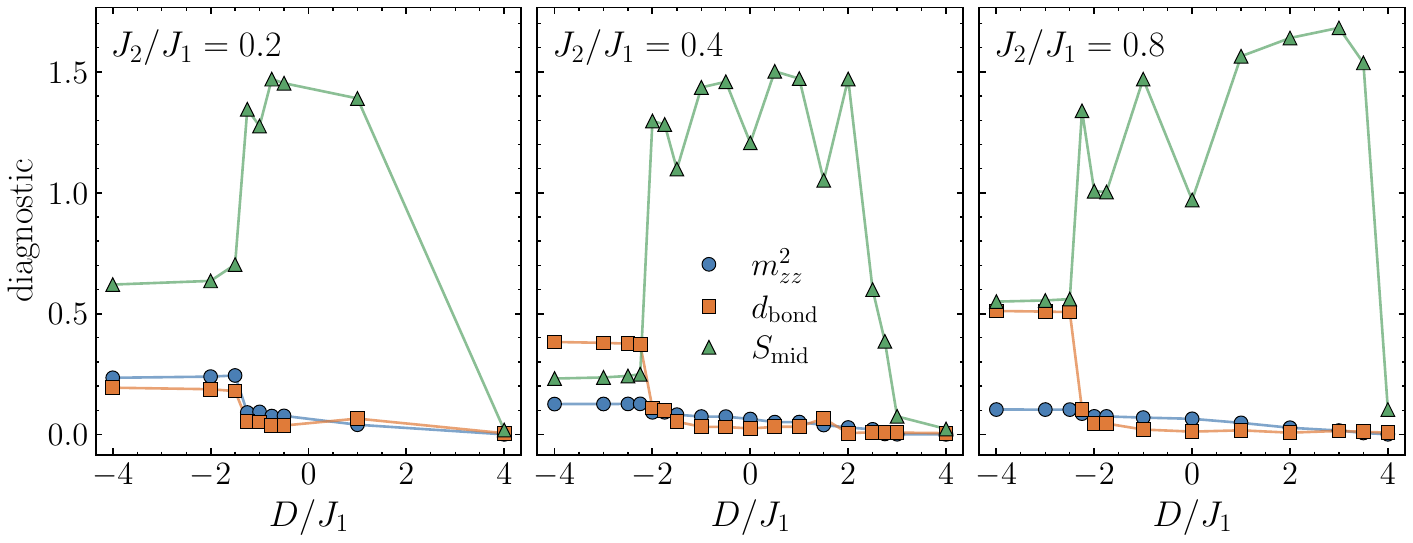}
    \caption{\textbf{Adaptive refinement of candidate phase boundaries.} Magnetic, bond, and entanglement diagnostics calculated by \sysname along adaptive cuts in $D$ at fixed $J_2=0.2$, $0.4$, and $0.8$ for systems of size $N=24$. Shown are the longitudinal structure-factor peak $m_{zz}^2$, the staggered bond-alternation proxy $d_{\rm bond}$, and the bipartite entanglement entropy $S_{\rm mid}$. Several of these quantities exhibit sharp changes over narrow ranges of $D$, with the locations of these features shifting toward more negative $D$ as $J_2$ is increased. At larger $J_2$, $d_{\rm bond}$ becomes increasingly pronounced on the negative-$D$ side, consistent with the candidate dimer region identified from finite-size scaling. The oscillations in $S_{\rm mid}$ at larger $D$ may reflect finite-size boundary effects, as open boundaries can induce Friedel-like oscillations in local observables \cite{lariviere2026}.}
    \label{fig:blbq_refinement_cuts}
\end{figure}

The benchmark Hamiltonians of the preceding sections have well-documented phase diagrams, making the agent's numerical results straightforward to validate, but also leaving open the possibility that the agent relies on prior knowledge rather than independently inferring the phase structure from its calculations. We therefore turn to a Hamiltonian whose complex two-dimensional phase diagram was published on the arXiv only recently (April 2026)~\cite{lariviere2026} and is unlikely to be represented in the training corpus of the considered backbones: the frustrated bilinear-biquadratic spin-1 \jj chain with single-ion anisotropy, given by
\begin{equation}
  H = \sum_{i}\Bigl[
      J_1\bigl(\cos\theta\,\mathbf{S}_i\!\cdot\!\mathbf{S}_{i+1}
             + \sin\theta\,(\mathbf{S}_i\!\cdot\!\mathbf{S}_{i+1})^2\bigr)
    + J_2\bigl(\cos\theta\,\mathbf{S}_i\!\cdot\!\mathbf{S}_{i+2}
             + \sin\theta\,(\mathbf{S}_i\!\cdot\!\mathbf{S}_{i+2})^2\bigr)
    + D\,(\hat{S}^z_i)^2 \Bigr],
  \label{eq:blbq_main_body}
\end{equation}
where $\mathbf{S}_i$ denotes a spin-1 operator at site $i$, $J_1$ and $J_2$ are the nearest- and next-nearest-neighbor exchange couplings respectively, and $D$ controls the single-ion anisotropy. We fix the bilinear-biquadratic angle at $\theta=0.2\pi$ and set $J_1=1$ as the energy scale, exploring $J_2\in[0,1]$ and $D\in[-4,4]$ with open boundary conditions (see Appendix~\ref{sec:hamiltonians} for further details).

The input specification handed to the agent contains the Hamiltonian, the ranges $J_2 \in [0,1]$ and $D \in [-4,4]$, and the instruction to treat the model as unknown (see input file in Appendix~\ref{sec:input_example}). The input file names no phase, states no symmetry, and suggests no observable, so every diagnostic below was chosen by the agent itself. We use GPT-5.6-Luna at high reasoning effort as the backbone and discuss a single run. We cap each Julia execution at 10 minutes and limit the total exploration time of \sysname to 8 hours. We note that these constraints are deliberately restrictive, in line with our goal of developing an efficient approach to help physicists build intuition about phase diagrams of unseen Hamiltonians.

\vspace{1ex}
\sysname checked the isolated two-site exchange (i.e.\ $N=2$, $J_2=0$, $D=0$) against its exact total-spin spectrum. At $\theta=0.2\pi$, the ground state is the triplet ($S^z_{\rm tot}=1$), with energy $E_0 = \sin\theta - \cos\theta = -0.221231742082474$. The agent reproduced this value to within $3\times10^{-17}$, and then repeated the same check with only the next-nearest-neighbor coupling active (i.e.\ $N=3$, $J_1=0$, $J_2=1$, $D=0$) to confirm that the $J_2$ bond connects sites $1$ and $3$ and that no nearest-neighbor term had been wrongly included. It wrote an independent exact-diagonalization code, cross-checking the results at $N=4$. It then numerically explored the symmetries of the Hamiltonian by computing commutator residuals for small systems, identifying total $S^z_{\rm tot}$, global $\pi$ rotations, time reversal and reflection as symmetries, while ruling out $S^x_{\rm tot}$ and $S^y_{\rm tot}$ at $D \neq 0$ and confirming full SU(2) at $D=0$. None of these symmetries were given in the specification. During validation, the agent identified and corrected two issues: 11 initial survey points had been computed in the wrong $S^z_{\rm tot}$ sector, and the connected quadrupole correlator contained an incorrect expansion. The affected results were excluded from the initial analysis, and the survey was subsequently regenerated. The initial DMRG survey covered $42$ points at $N=24$ with a maximum bond dimension $384$, followed by adaptive cuts in $D$ at five values of $J_2$ and size ladders reaching $N=60$. In total, the agent performed $222$ DMRG calculations spanning $158$ distinct $(J_2, D, N)$ parameter points, which cover $83$ distinct $(J_2, D)$ coordinates of the phase diagram at system sizes $N = 4$, $12$, $24$, $36$, $48$ and $60$, supplemented by exact diagonalization at $N = 4$, $6$ and $8$.

Having validated its numerical implementation, \sysname proceeds to explore the phase diagram by selecting and calculating a range of magnetic, bond, nonlocal and entanglement diagnostics, conducting finite-size extrapolations where it deems necessary. From these results, the agent identifies five candidate regions, summarized in Table~\ref{tab:blbq_found} and shown in Fig.~\ref{fig:blbq_phase_diagram_compare}. We discuss the evidence it uncovers for each candidate below.

We begin with the Ising phase, which the agent identifies in the $D\sim[-4,-1]$, small $J_2$ region of the phase diagram primarily by considering the static spin structure factor
\begin{equation} \label{eq:structure_factor}
    S_{\alpha\alpha}(q) = \frac{1}{N^2}\sum_{j,k}\langle \hat{S}^\alpha_j \hat{S}^\alpha_k\rangle e^{iq(j-k)}
\end{equation}
for both $\alpha=z$ and $\alpha=x$. In the candidate Ising-ordered region, the longitudinal structure factor $S_{zz}(q)$ peaks at $q=q^*=\pi$, consistent with staggered, period-two spin correlations $\langle \hat{S}^z_j \hat{S}^z_{j+r}\rangle$ as a function of the separation $r$. We denote the corresponding peak values by
\begin{equation} \label{eq:structure_factor_peak}
m_{\alpha\alpha}^2 \equiv \max_q S_{\alpha\alpha}(q),
\end{equation}
which provide finite-size diagnostics of magnetic order. The agent measures the scaling of both the transverse peak $m_{xx}^2$ and the longitudinal peak $m_{zz}^2$, and finds that the former decays approximately as $1/N$, while the latter remains between $0.38$ and $0.41$ from $N=24$ to $N=60$ at $(J_2,D)=(0,-2)$ and $(0,-4)$. At $(J_2,D)=(0,-1)$, where the ladder reaches $N=48$, a linear extrapolation in $1/N$ yields a nonzero intercept of $m_{zz}^2\approx0.431$. The same extrapolation at $(0.2,-2)$ gives $0.113$, an order-parameter amplitude roughly four times smaller, so the agent's N\'eel assignment there rests on a weaker signal than at $J_2=0$. This is strong finite-size evidence for long-range longitudinal spin order, and that, coupled with the staggered wavevector $q^*=\pi$, supports the agent's identification of a N\'eel-ordered Ising antiferromagnet candidate.

Next, \sysname finds strong evidence of dimer order at several points in the large-$J_2$, $D\sim[-4,-2]$ region, identifying a candidate dimer phase analogous to the bond-ordered phase of the spin-$1/2$ \jj chain. In particular, it measures the staggered bond alternation
\begin{equation}
    \label{eq:bond_alternation}
    d_\text{bond} = \frac{1}{N-1} \left| \sum_{i=1}^{N-1} (-1)^i \langle\mathbf{S}_i\cdot\mathbf{S}_{i+1}\rangle\right|.\qquad
\end{equation}
Here $d_\text{bond}$ serves as a finite-size dimer order parameter, probing the alternation between strong and weak nearest-neighbor bonds, as expected in a state with enhanced singlet correlations on every other bond. At a representative point within this candidate region, $(J_2,D)=(0.4,-4)$, the agent finds that $d_\text{bond}$ decreases only from $0.383$ at $N=24$ to $0.340$ at $N=60$, with a linear extrapolation in $1/N$ yielding a nonzero intercept of approximately $0.313$. This persistence of the bond alternation provides strong finite-size evidence for long-range dimer order in the thermodynamic limit. The identification is further supported by the connected bond structure factor $S_c^{BB}(q)$, analogous to the spin structure factor in Eq.~\eqref{eq:structure_factor}, but this time constructed from connected correlations between nearest-neighbor bond operators rather than unconnected spin operators. At $N=24$, the agent finds $S_c^{BB}(\pi)=0.0441$ at $(J_2,D)=(0.4,-4)$, compared with $0.0153$ at a representative point $(J_2,D)=(0,0)$ in the proposed Haldane phase, to be discussed next. This enhanced staggered bond-correlation signal provides additional support for the proposed dimer order.

Something interesting happens when \sysname considers the boundaries of these candidate regions. In Fig.~\ref{fig:blbq_refinement_cuts}, the agent performs adaptive cuts in $D$ at fixed $J_2$ for $J_2=0.2$, $0.4$ and $0.8$, covering a range of frustration, for systems of size $N=24$. Several of the calculated diagnostics exhibit sharp changes as $D$ increases, providing finite-size evidence for possible phase transitions. The character of these cuts also evolves with $J_2$. At smaller $J_2$, $m_{zz}^2$ is substantially larger on the negative-$D$ side, consistent with the magnetic order discussed above. As $J_2$ increases, this magnetic signal weakens, while the bond-alternation proxy becomes increasingly pronounced in the $D\sim[-4,-2]$ region, consistent with the emergence of the candidate dimer region. Upon increasing $D$, the bond-alternation proxy drops sharply, accompanied by yet another sharp change in the entanglement entropy across the central bond, $S_{\rm mid}$. Interestingly, at larger $J_2$, these abrupt changes are qualitatively consistent with the first-order dimerized--Floating-2 transition found in Ref.~\cite{lariviere2026}, although the agent does not obtain sufficient evidence to determine the transition order. Just as importantly, the locations of these sharp features shift toward more negative $D$ as $J_2$ is increased. The agent brackets them at $D\in[-1.50,-1.25]$ for $J_2=0.2$, $D\in[-2.00,-1.75]$ for $J_2=0.4$, and $D\in[-2.50,-2.25]$ for $J_2=0.8$. Although the agent initially interprets these brackets as marking a dimer-to-non-dimer crossover, its eventual identification of a candidate N\'eel-ordered point at $(J_2,D)=(0.2,-2)$ challenges this interpretation at small $J_2$, while its finite-size scaling supports persistent dimer order at larger $J_2$. We therefore use the systematic shift of these features, rather than their finite-$N$ phase assignments, to motivate the negative slope of the corresponding boundary in our schematic interpolation in Fig.~\ref{fig:blbq_phase_diagram_compare}. The ability to infer this boundary geometry from adaptive calculations represents an important milestone for \sysname.

As $D$ increases towards zero and enters the positive regime, the agent uncovers signatures of a symmetry-protected topological Haldane phase. Its primary evidence comes from the nonlocal longitudinal string correlator
\begin{equation}
    \label{eq:string_correlator}
    \mathcal{O}_{\rm str}(i,j)
    =
    \left\langle
    \hat{S}_i^z
    \left(\prod_{i<k<j} e^{i\pi \hat{S}_k^z}\right)
    \hat{S}_j^z
    \right\rangle.
\end{equation}
At $(J_2,D)=(0,0)$, the agent finds that the string correlator, evaluated at the largest available separation, increases in magnitude from $0.144$ at $N=24$ to $0.185$ at $N=60$, with a linear extrapolation in $1/N$ producing $|\mathcal{O}_{\rm str}(i,j)|\approx0.206$. These surviving nonlocal string correlations are signatures of possible Haldane order. Along the $D=0$ line, however, $|\mathcal{O}_{\rm str}(i,j)|$ for $N=48$ decreases with increasing $J_2$, falling from $0.0450$ at $J_2=0.4$ to $0.0084$ at $J_2=1$. This suggests a possible transition out of the proposed Haldane phase into a frustrated non-Haldane region when $J_2$ is large enough, and the agent proposes a boundary around $J_2 \sim 0.5$, as can be seen in Fig.~\ref{fig:blbq_phase_diagram_compare}.

\begin{table}[tbh]
  \centering\small
  \caption{The five candidate regions the agent reports for Eq.~\eqref{eq:blbq_main_body}
  at $\theta=0.2\pi$, representative labelled points or ranges, the finite-size evidence it cites for each, and the corresponding phase of Ref.~\cite{lariviere2026}. The last row is a reference phase it does not report.}
  \label{tab:blbq_found}
  \renewcommand{\arraystretch}{1.25}
  \begin{tabular}{@{}p{0.16\linewidth}p{0.19\linewidth}p{0.38\linewidth}p{0.18\linewidth}@{}}
    \toprule
    \textbf{Agent's label} & \textbf{Labelled points} & \textbf{Evidence cited by the agent} & \textbf{Ref.~\cite{lariviere2026}} \\
    \midrule
    N\'eel, Ising-ordered &
      $(0,-1)$, $(0,-2)$, $(0,-4)$, $(0.2,-2)$ &
      Staggered $m^2_{zz}$ survives $1/N$ extrapolation, with intercept $0.431$
      at $(0,-1)$ and a smaller but nonzero $0.113$ at $(0.2,-2)$. At $(0,-2)$
      and $(0,-4)$ it holds between $0.38$ and $0.41$ up to $N=60$, while the
      transverse peak decays as $1/N$ &
      Ising-AFM \\
    Haldane, SPT &
      $(0,0)$, $(0,1)$, $(0,2)$ &
      String correlator moves from $-0.144$ ($N=24$) to $-0.185$ ($N=60$), intercept $-0.206$, edge
      splitting falls $0.0207 \to 0.0071$ between $N=36$ and $48$ &
      Haldane \\
    Dimer, bond-order &
      $J_2 \geq 0.4$ at $D=-4$, $(0.6,-2.5)$, $(1,-2.5)$ &
      Bond alternation $0.383 \to 0.340$ from $N=24$ to $60$, intercept $0.313$, connected bond structure factor at $q=\pi$ of $0.0441$, against $0.0153$ in
      the Haldane candidate &
      dimerized \\
    Trivial large-$D$ &
      $(0,3)$, $(0,4)$, $(0.4,3)$, $(0.4,4)$ &
      $m^2_{zz}$ falls from $4.05\times10^{-4}$ to $1.68\times10^{-4}$ between $N=24$ and $60$; leading Schmidt weight = $0.9984$, entanglement entropy = $0.0234$, and magnetic gap stable at $\Delta E\simeq1.25$ &
      large-$D$ \\
    Incommensurate correlations &
      $(0.4,0)$, $(0.8,0)$, $(1,0)$, $(0.8,4)$, $(1,4)$ &
      Magnetic peaks decrease with size; $q^*$ varies across the frustrated region; enhanced entanglement entropy; string-correlator fit at $(0.6,0)$ gives an intercept of $-0.0087$, consistent with zero. Gaplessness remains unestablished &
      Floating and trimerized regions; partial overlap with Haldane \\
    \midrule
     Not identified & --- & No period-three order parameter constructed, despite measuring $q^*\approx2\pi/3$ along $D=0$ & trimerized \\
    \bottomrule
  \end{tabular}
\end{table}

At large and positive $D$, the agent independently identifies a candidate topologically trivial large-$D$ phase. It calculates the onsite expectation value $\langle (\hat{S}_i^z)^2\rangle$ and finds it strongly suppressed, consistent with a wavefunction dominated by the product state $|0,0,\ldots,0\rangle$. This picture is further supported by a Schmidt decomposition across the central bond yielding a leading Schmidt value of $0.9984$ as well as a very small bipartite entanglement entropy of $S_{\rm mid}=0.0234$. The absence of magnetic order is equally apparent. For example, at $(J_2,D)=(0,4)$, the agent finds that $m_{zz}^2$ decreases from $4.05\times10^{-4}$ at $N=24$ to $1.68\times10^{-4}$ at $N=60$, extrapolating to $1.2\times 10^{-5}$ as $N\to\infty$. Finally, the magnetic gap $\Delta E=E_{M=1}-E_{M=0}$ remains nearly constant, changing from $1.248$ at $N=24$ to $1.250$ at $N=36$, in contrast to the decreasing edge gap observed at candidate Haldane points.

The agent seeks next to characterize the frustrated region beyond the suspected Haldane phase. Firstly, it notices that the normalized structure-factor peak $m_{zz}^2$ decreases with system size, falling from $0.0638$ at $N=24$ to $0.0365$ at $N=48$ for $(J_2,D)=(0.4,0)$. It further observes that for chain lengths up to $N=60$, the dominant wavevector $q^*$ of the longitudinal structure factor varies across the frustrated region, particularly for $D\sim[-2,2]$ and $J_2\gtrsim0.5$, rather than remaining fixed at the N\'eel value $q^*=\pi$, as shown in Fig.~\ref{fig:blbq_phase_diagram_points}. This leads the agent to suspect incommensurate correlations, which is generally not enough to establish a distinct phase. Interestingly, it also measures $q^*\approx2.086$ at $(J_2,D)=(0.4,0)$, without recognizing its proximity to the commensurate period-three wavevector $2\pi/3$ that could be indicative of trimer order (more on this below). We note that, whereas the agent extracts $q^*$ from the peak of the longitudinal spin structure factor, Ref.~\cite{lariviere2026} determines the characteristic wavevector by fitting Friedel oscillations in local correlation functions. Both diagnostics probe characteristic spatial modulations, and their extracted wavevectors are therefore expected to be related, although they need not coincide exactly, particularly for finite open chains.

The suppression of the nonlocal string correlator along the $D=0$ line with increasing $J_2$ leads the agent to suspect a "distinct correlation regime". Notably, it independently raises the possibility that this regime may be gapless, but recognizes that its evidence is insufficient to establish this, explicitly identifying the lack of several necessary diagnostics, including a reliable central-charge fit and bulk-gap extrapolation. Thus, it refrains from assigning the gapless label.

The agent leaves a number of sampled points unresolved throughout the phase diagram. In the upper-right portion, several sampled points remain without a confident phase assignment based on the evidence gathered by the agent, leading us to indicate an unresolved region in Fig.~\ref{fig:blbq_phase_diagram_compare}. We emphasize that we are not claiming that the agent's proposed phase assignments are definitive identifications $-$ they are only candidates. Its DMRG calculations provide finite-size evidence consistent with the emergence of the corresponding phases in the thermodynamic limit, but further numerical investigation is required to establish their existence conclusively.

Compared to the reference phase diagram of Ref.~\cite{lariviere2026}, an adaptation of which is depicted in Fig.~\ref{fig:blbq_phase_diagram_compare}, the most glaring omission on the part of \sysname is the failure to uncover signatures of the gapped, trimerized phase, instead grouping much of the frustrated region under its incommensurate-correlation candidate. Furthermore, it does not reliably characterize the phase boundaries or determine their transition types, some of which are themselves challenging to resolve numerically~\cite{lariviere2026}. Nevertheless, it is encouraging that the agent independently raises the possibility of gapless behavior and explicitly identifies bulk-gap extrapolation as a missing diagnostic in this regard. Such a calculation could help distinguish the gapless floating regions from the intervening gapped trimerized phase, while additional period-three bond-order measurements, motivated by physical intuition about frustrated spin chains, would be needed to establish the latter's symmetry breaking \cite{corboz2007spontaneous,lariviere2026}. This illustrates a potential role for human-in-the-loop interventions: drawing on prior physical knowledge to guide the agent towards meaningful, targeted calculations when its existing diagnostics leave competing phase interpretations unresolved. We also firmly believe that further improvements can be realized by increasing the computational time budget and refining the engineering underpinning the agent framework (e.g. skills, tools, etc.). Indeed, 8 hours of total runtime and 10 minutes per Julia call were restrictive choices, but deliberately so $-$ the agent nevertheless manages to produce a surprisingly detailed phase diagram that we find useful for developing intuition about the physics of an unfamiliar Hamiltonian.

\section{Conclusion and outlook}

We have introduced \sysname, an agentic toolbox that explores the phase diagram of strongly-correlated quantum spin Hamiltonians. Given the Hamiltonian, it relies on DMRG and exact-diagonalisation simulations to unravel the phase diagram. On benchmarks with known answers, it locates the transitions, and on the likely unseen frustrated bilinear-biquadratic spin-1 \jj chain with anisotropy, it recovers most of the phases found in Ref.~\cite{lariviere2026} $-$ including phases absent from the related model studied years earlier in Ref.~\cite{corboz2007spontaneous} $-$ places them correctly in the plane, and begins to capture the effect of frustration on the candidate phase boundaries, all at a cost of less than 10 dollars per study. The method benefits heavily from a set of curated knowledge skill sets. A compact skill describing good computational practices anticipates a significant number of possible code failures when using research-grade code bases. With this skill set, open-weight models served locally can be made competitive with cloud-hosted ones on these tasks.

Our work uses DMRG and ED, due to their reliability and robustness. However, extending \sysname to other computational methods could allow it to study Hamiltonians beyond the reach of matrix product state methods in the future, whether through variational Monte Carlo (VMC) and neural quantum states in two and three dimensions~\cite{carleo2017nqs,vicentini2022netket}, projected entangled pair states~\cite{verstraete2008peps,orus2014}, unbiased quantum Monte Carlo (QMC) ~\cite{sandvik2010lectures,assaad2022alf}, or dynamical mean-field theory for multi-orbital systems~\cite{georges1996dmft,rohringer2018diagrammatic,parcollet2015triqs}. Nevertheless, every method comes with its own limitations, from variational bias in VMC \cite{rigo2026neural} to the sign problem and the ill-conditioned analytic continuation of imaginary-time data in QMC~\cite{troyer2005sign,jarrell1996maxent}, all of which would need to be tackled in the future to enable reliable autonomous exploration with these methods. On top of opening up the range of Hamiltonians that can be studied, a further benefit of incorporating multiple computational methods is the possibility of cross-validation between them. The DMRG-against-ED cross-check presented here provides a simple example: agreement between methods with independent failure modes increases confidence in the numerical results, and more generally can provide stronger evidence for a candidate phase than either method alone (see, e.g., the multi-method analysis in Ref.~\cite{qin2020absence}).

A second extension concerns what the agent measures. A decade of work has automated the phase classification step itself, through supervised, confusion-based and unsupervised detection of order directly from sampled configurations~\cite{carrasquilla2017machine,vannieuwenburg2017confusion,wetzel2017unsupervised,carrasquilla2020review, zhou2026autonomousphasediscovery}. We do not use these methods here, and the agent instead selects conventional observables, but exposing them as an additional tool would allow it to search for order even when it fails to identify an appropriate order parameter.

Beyond expanding its computational and diagnostic capabilities, the benefits of expert skills suggest a route towards self-improvement through an evolving, human-readable library of analysis strategies, building on recent work on adaptive contexts and reusable skills~\cite{zhang2025ace,zhou2026mementoskills}. After each study, \sysname could propose explicit revisions to this library, documenting the limitation encountered, the supporting numerical evidence, and how future analyses should change. Subjecting such revisions to expert review and validation on independent benchmark Hamiltonians would provide a testable route to self-improvement while keeping the evolving analysis procedures accessible to humans.

Finally, we must stress our vision for \sysname and agent-based frameworks of its kind. While we firmly believe that there is significant room for growth in both skill development and the engineering design of our tool, we do not consider it a replacement for the human scientist. On the contrary, we envision a future in which tools like \sysname work hand-in-hand with human researchers, and our results provide a first glimpse as to how such a workflow might operate. The phase diagram in Fig.~\ref{fig:blbq_phase_diagram_compare}, for example, provides substantial physical intuition, but it clearly does not capture every feature of the underlying problem. This is precisely where the human can re-enter the loop. The appearance of correlations near $q=2\pi/3$, for instance, may prompt a researcher familiar with spin-$1$ physics to ask whether the agent has overlooked trimerization, and to request specific measurements of period-three order. Likewise, physical intuition about the possible nature of a transition can motivate further exploration. For example, a researcher may suspect a BKT transition along one of the candidate boundaries and direct the agent toward the diagnostics needed to test this hypothesis $-$ a suspicion borne out in the phase diagram of Ref.~\cite{lariviere2026}. The resulting picture is one of human-in-the-loop scientific discovery: not one in which the agent replaces the scientist, but one in which the scientist leverages the agent as a tool and works closely with it. The agent explores broadly and autonomously, the human recognizes what is interesting or missing, and the agent returns to the problem with a more targeted objective.

\section*{Code availability}
The code is made public on GitHub: \url{https://github.com/jwnys/quantum_matter_explorer.git}.

\hfill
\newpage

\appendix
\section{Skills and tools}\label{app:skillstools}

\begin{table}[h!]
  \centering\small
  \caption{The skill library available to the agent, grouped by the stage of the study each serves: setting up and validating the model and coding, exploring and identifying phases, and reporting. A short description of every skill is always present in context, and the full text is read on demand when
  the agent judges it relevant.}
  \renewcommand{\arraystretch}{1.25}
  \begin{tabular}{@{}p{0.22\linewidth}p{0.72\linewidth}@{}}
    \toprule
    \textbf{Skill} & \textbf{Role} \\
    \midrule
    \tool{model-setup} & Read before any Hamiltonian code is written. Turns a written
      input specification into a Hamiltonian that can be trusted, covering which operator algebra the
      spec implies, which conventions are assumed, and how
      to choose system sizes commensurate with the periodicities the Hamiltonian could break. \\
    \tool{ed-validation} & Read before writing any exact-diagonalisation script, and whenever a validation check fails. Covers how to prove that the DMRG code implements the intended model,
      by reproducing an analytically known limit and by cross-checking a small chain against an
      independent ED calculation. \\
    \tool{itensor-dmrg} & Read before writing or editing any ITensor Julia file. A code cookbook for
      the Hamiltonian, the observables and the parameterised scan script, covering the pitfalls of
      the libraries. This is the skill whose ablation is reported
      in Fig.~\ref{fig:fig9d_julia_skill_effort}. \\
    \midrule
    \tool{observables} & Read before Stage~1 of the survey, and again whenever a region cannot be classified. Provides expert guidance on what to measure for a Hamiltonian whose phases are not known in advance, including reasoning from the symmetries of the model. \\
    \tool{transitions} & Read in Stage~2, when the survey has shown where something changes. Covers bisecting a boundary in a window rather
      than refining the whole range, and which quantities or estimators can potentially locate a transition without assuming its character. \\
    \tool{phase-id} & Read once a phase label is needed. A
      decision tree from measured observables to a label, separating long-range order, gapless behavior, symmetry-protected topological order and trivial gapped phases, and stating when the honest outcome is that the region remains unresolved. \\
    \tool{scaling} & Read in Stage~4, and before quoting any number that describes the infinite
      system, such as an order parameter, a gap, a central charge or a transition location. Covers which quantity to extrapolate, and pitfalls related to finite-size effects. \\
    \tool{disorder} & Covers hints about how many realisations are needed in disordered Hamiltonians, and which
      quantities must be averaged. \\
    \midrule
    \tool{reporting} & Read when writing up, and before the study is declared finished. Covers what
      the research note must contain. \\
    \bottomrule
  \end{tabular}
  \label{tab:skills}
\end{table}

\begin{table}[h!]
  \centering\small
    \caption{The tools available to the deep agent.}
  \renewcommand{\arraystretch}{1.25}
  \begin{tabular}{@{}p{0.22\linewidth}p{0.72\linewidth}@{}}
    \toprule
    \textbf{Tool} & \textbf{Role} \\
    \midrule
    \tool{run\_julia} & Runs a Julia command, used, e.g., to run a single DMRG calculation through the ITensorMPS library. Returns the output of the Julia script, which typically includes principal convergence diagnostics --- the truncation error, the total $S^z$, and the final energy --- so that the reliability of the result can be judged immediately. \\
    \tool{run\_julia\_batch} & Similar to \tool{run\_julia}, but can run a script with multiple inputs in parallel. \\
    \tool{validate} & Performs a fast syntax check of a Julia or Python script. \\
    \tool{lookup\_doc} & Retrieves the exact call signature and keyword arguments of a library function from a local mirror of e.g.\ the ITensor source code and documentation, a retrieval-augmented step~\cite{lewis2020rag}. This lets the agent call an unfamiliar routine correctly, instead of guessing its interface from memory and risking a silent error. \\
    \tool{update\_run\_summary} & Appends a concise entry to a running summary file that acts as the study's human-readable record. \\
    \tool{finish} & Signals that the study is complete and brings the run to a stop. \\
    \midrule
    \tool{request\_plan\_review} & Pauses after \code{plan.md} is drafted, before any full simulation, so a human can review and edit the plan. \\
    \tool{request\_review} & Pauses once the study looks complete, before \tool{finish}, so a human can review the final note and possibly trigger additional simulations. \\
    \midrule
    \tool{write\_todos} & Maintains a frequently updated checklist of the sub-tasks the agent is working through. The durable research plan itself is kept separately, in \code{plan.md}, so that short-term bookkeeping does not crowd out the longer-term strategy. \\
    \tool{read\_file}, \tool{write\_file}, \tool{edit\_file}, \tool{ls}, \tool{glob}, \tool{grep} & Provide file-system operations.\\
    \tool{execute} & Runs shell commands inside the run directory, used for e.g.\ lightweight numerical post-processing in Python. \\
    \tool{task} & Delegates a narrowly scoped sub-task to one of the isolated subagents (general-purpose, data-analyst or physics-critic). The subagent works in its own separate context and returns only a compact result, which keeps bulky intermediate reasoning out of the main agent's history. \\
    \tool{compact\_conversation} & Compresses the conversation history, freeing up context~\cite{lodha2026lesscontext,packer2023memgpt}. \\
    \bottomrule
  \end{tabular}
  \label{tab:tools}
\end{table}

\section{Reference Hamiltonians}\label{sec:hamiltonians}
We list the Hamiltonians used in the results section.

\subsection{Anisotropic XY chain}
First, the anisotropic XY chain in a transverse field
\begin{equation}
  H = -\sum_{i=1}^{L-1}\left(\frac{1+\gamma}{2}\,\hat{X}_i\hat{X}_{i+1}
                                +\frac{1-\gamma}{2}\,\hat{Y}_i\hat{Y}_{i+1}\right)
          - \frac{h}{2}\sum_{i=1}^{L} \hat{Z}_i .
  \label{eq:xy}
\end{equation}
This model is exactly solvable through a Jordan--Wigner transformation onto free fermions for all $(\gamma,h)$~\cite{lieb1961,barouch1971}. For any $\gamma \neq 0$, the model undergoes an Ising quantum phase transition at $h_c=1$, with central charge $c=\tfrac12$~\cite{pfeuty1970}. In the isotropic limit $\gamma=0$, the model reduces to the XX chain, which instead exhibits a transition from a gapless Luttinger liquid ($c=1$) to a fully polarised phase at $h_c=1$.

\subsection{Frustrated $J_1$--$J_2$ chain}

The frustrated $J_1$--$J_2$ chain is given by
\begin{equation}
  H = J_1\sum_{i}\mathbf{S}_i\cdot\mathbf{S}_{i+1}
        + J_2\sum_{i}\mathbf{S}_i\cdot\mathbf{S}_{i+2},
  \qquad \alpha \equiv J_2/J_1 ,
  \label{eq:j1j2}
\end{equation}
with $J_1>0$ and $J_2\geq 0$.

We study the dimerization transition at $\alpha_c = 0.2411(1)$~\cite{okamoto1992}, a Berezinskii--Kosterlitz--Thouless (BKT) transition from a critical Luttinger liquid to a gapped, spontaneously dimerized phase, and the Lifshitz point at $\alpha = 0.52036(6)$~\cite{deschner2013}. At the exactly solvable Majumdar--Ghosh point, $\alpha = 1/2$, the ground state consists of a product of nearest-neighbor singlets, with an exact energy of $-3/8$ per site for periodic chains of even length~\cite{majumdar1969}.

\subsection{Frustrated bilinear-biquadratic spin-1 chain}

The frustrated bilinear-biquadratic spin-1 chain adds a single-ion anisotropy to a next-nearest-neighbor coupling,
\begin{equation}
  H = \sum_{i}\Bigl[
      J_1\bigl(\cos\theta\,\mathbf{S}_i\!\cdot\!\mathbf{S}_{i+1}
             + \sin\theta\,(\mathbf{S}_i\!\cdot\!\mathbf{S}_{i+1})^2\bigr)
    + J_2\bigl(\cos\theta\,\mathbf{S}_i\!\cdot\!\mathbf{S}_{i+2}
             + \sin\theta\,(\mathbf{S}_i\!\cdot\!\mathbf{S}_{i+2})^2\bigr)
    + D\,(\hat{S}^z_i)^2 \Bigr],
  \label{eq:blbq}
\end{equation}
where $\mathbf{S}_i$ are spin-1 operators and the single angle $\theta$ fixes the ratio of bilinear to biquadratic exchange on both bonds. We set $J_1=1$ and $\theta=0.2\pi$, away from the SU(3)-symmetric line, and explore the $(J_2,D)$ plane. The model was studied in Ref.~\cite{corboz2007spontaneous} with $D=0$. La Rivi\`{e}re and Chepiga~\cite{lariviere2026} report five gapped phases, namely Haldane, large-$D$, Ising antiferromagnetic, trimerized and dimerized, separated by Gaussian, Ising and Kosterlitz-Thouless transitions, and in addition two incommensurate critical floating phases. The floating phases were resolved with bond dimensions up to $4000$, roughly an order of magnitude beyond what the agent uses here, so they are a deliberately demanding target rather than an expected one.

Both of these references differ from those mentioned in relation to Eqs.~\eqref{eq:xy} and~\eqref{eq:j1j2}. They report numerical phase diagrams rather than exact solutions, obtained by the original authors with the same method the agent uses.

\section{Input example}\label{sec:input_example}
Below, we provide an example for the input file specifying the
frustrated spin-$1/2$ chain in Eq.~\eqref{eq:j1j2}.

\begin{lstlisting}
\documentclass{article}
\usepackage{amsmath}
\begin{document}

\section*{Frustrated $J_1$-$J_2$ Heisenberg Chain}

\begin{equation}
H = J_1 \sum_{i} \mathbf{S}_i \cdot \mathbf{S}_{i+1}
  + J_2 \sum_{i} \mathbf{S}_i \cdot \mathbf{S}_{i+2}
\end{equation}

where $J_1 > 0$ is the nearest-neighbour antiferromagnetic coupling and
$J_2 \geq 0$ is the next-nearest-neighbour coupling (frustration parameter).
We assume periodic boundary conditions and a chain of length $L$.
Set $J_1 = 1$.

The frustration ratio $\alpha = J_2/J_1$ is the key parameter.
Explore $\alpha \in [0, 1]$.

\end{document}
\end{lstlisting}

Similarly, for the example of the spin-1 chain in Eq.~\eqref{eq:blbq}, we input:
\begin{lstlisting}
# Frustrated Bilinear-Biquadratic Spin-1 Chain with Single-Ion Anisotropy

## Hamiltonian

$$
H = \sum_{i}\Bigl[
  J_{1}\bigl(\cos\theta \mathbf{S}_{i} \cdot \mathbf{S}_{i+1}
           + \sin\theta (\mathbf{S}_{i} \cdot \mathbf{S}_{i+1})^{2}\bigr)
+
  J_{2}\bigl(\cos\theta \mathbf{S}_{i} \cdot \mathbf{S}_{i+2}
           + \sin\theta (\mathbf{S}_{i} \cdot \mathbf{S}_{i+2})^{2}\bigr)
+ D(S^{z}_{i})^{2} \Bigr]
$$

where:

- $\mathbf{S}_i$ are **spin-1** operators ($S = 1$, so $S^{z}$ has eigenvalues $-1, 0, +1$).
- Each bond carries both a **bilinear** term $\mathbf{S}_i\cdot\mathbf{S}_j$ and a **biquadratic**
  term $(\mathbf{S}_i\cdot\mathbf{S}_j)^2$, mixed by the single angle $\theta$.
- $J_1$ is the nearest-neighbour coupling and $J_2$ the next-nearest-neighbour coupling.
- $D$ is a **single-ion anisotropy** acting on every site.
- Open boundaries.

Fix $J_{1} = 1$ to set the energy scale, and fix the bilinear-biquadratic angle at

$$\theta = 0.2\pi .$$

## Parameter ranges to explore

- $J_{2} \in [0, 1]$
- $D \in [-4, 4]$

## Notes

Treat as **unknown**. Discover the phases and the boundaries between them from scratch. Do not
assume how many phases there are.
\end{lstlisting}

\section{Time consumption}\label{sec:timeconsumption}
We report the wall-clock time in terms of the subcomponents of the tool: LLM usage, Julia runs, and other tools. We observe that close to 50\% of the time is spent on LLM inference for local models, which is lower for cloud-based models. The experiments are each run on two H100s, and the Julia calculations run on CPUs, with a maximum of 8 simulations run in parallel. Qwen spends the most wall-clock time on inference (61 \%) and the least on simulation (38\%), while GPT-5.6-luna (at low reasoning effort) inverts both (29\% / 70\%) in a third of the total time.

\begin{figure}[htb]
    \centering
    \includegraphics[width=0.5\linewidth]{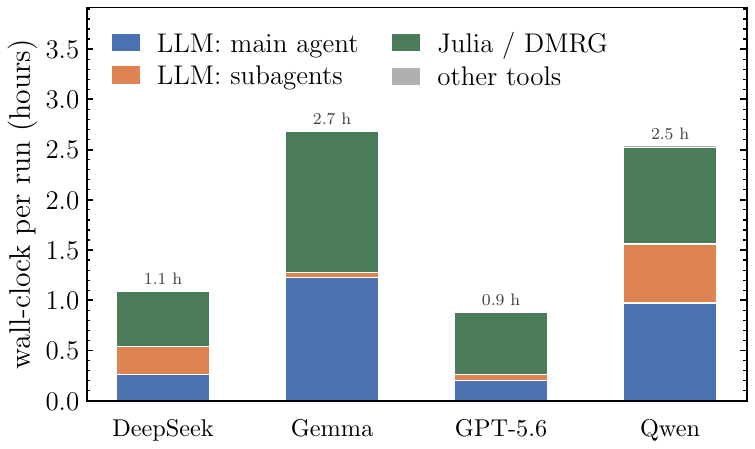}
    \caption{\textbf{Wall-clock time per run}. Mean time consumed per run for various backbones, separated per category: LLM, Julia runs, and other tools. Results are averaged over two Hamiltonians (XY and \jj model), with four repeated runs for each.}
    \label{fig:fig7b_time_split}
\end{figure}

The cloud backbone was run on OpenAI's flex service tier, priced at USD 0.10 per million input tokens, USD 0.01 per million cached input tokens, and USD 0.60 per million output tokens. A single study consumed on average $1.0\times10^{7}$ input and $6.9\times10^{4}$ output tokens over 163 model calls, of which at most 92\% of the input was eligible for prefix caching. No request exceeded the 272{,}000-token threshold above which input is billed at twice the rate. Billing every input token at the uncached rate gives an upper bound of USD 1.05 per study, and billing every repeated prefix at the cached rate gives a lower bound of USD 0.23 per run.

\section{Agent-generated research note}
\label{app:agent_output}

The research note generated by \sysname for the spin-1 chain of Eq.~\eqref{eq:blbq_main_body} is reproduced in its entirety at the end of this manuscript, following the references. It is important to state that the agent does not include in the note every calculation it performs during the exploration. For example, the phase diagram presented in the note displays only a subset of the computed parameter points, whereas Fig.~\ref{fig:blbq_phase_diagram_compare} in the main body of this article includes all the additional points calculated by the agent.

\clearpage
\bibliographystyle{apsrev4-2}
\bibliography{review}

\clearpage
\includepdf[
    pages=-,
    pagecommand={}
]{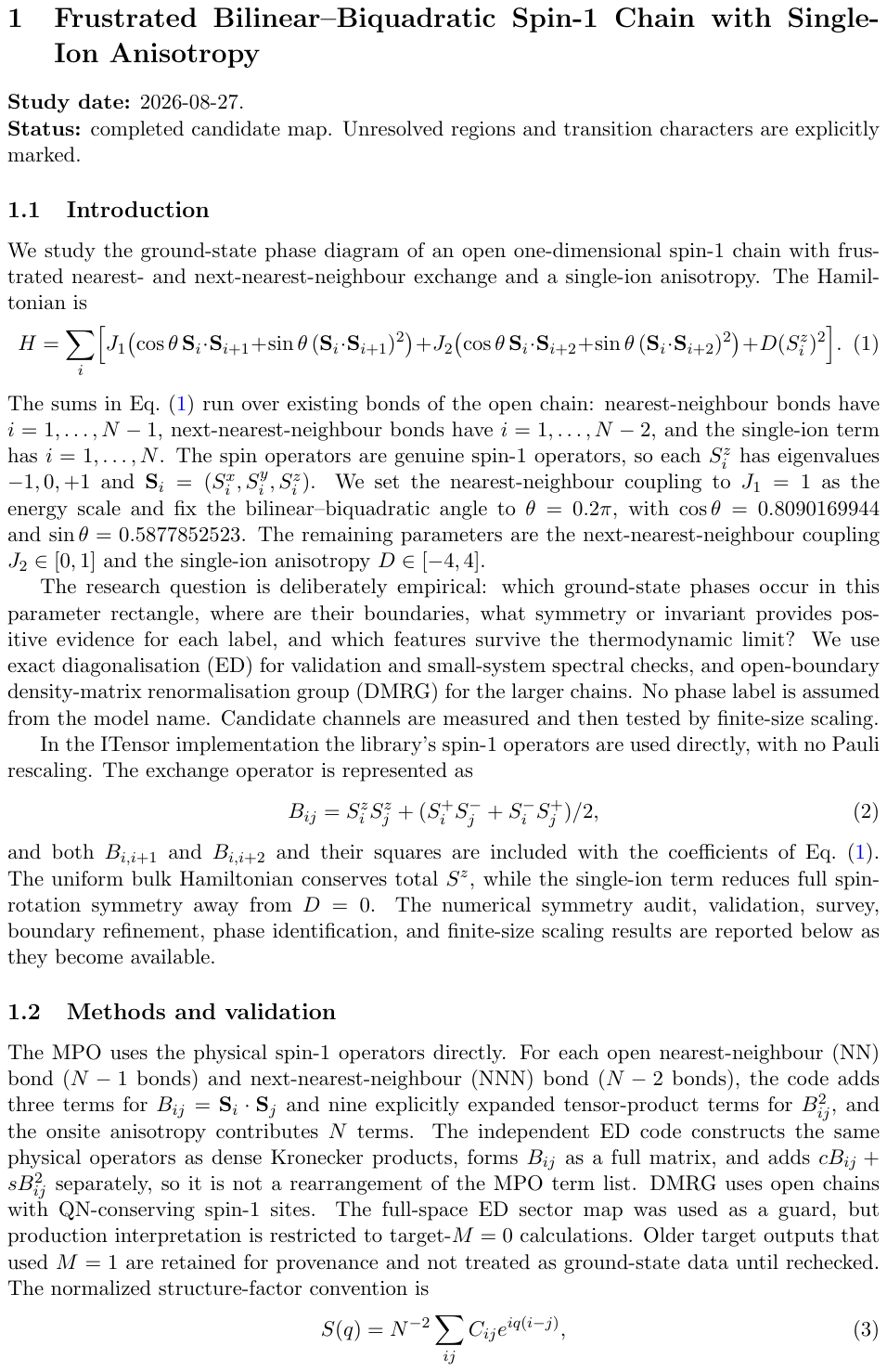}

\end{document}